\documentclass[journal]{IEEEtran}

\usepackage{graphicx} \usepackage{bm} \usepackage{amsmath}
\usepackage{cite}
\usepackage{amssymb} 

\usepackage{amsmath,amsthm,amssymb,mathrsfs}
\usepackage{bm}

\ifCLASSINFOpdf
\else
\fi
\begin{document}

\title{Coupled dynamics of magnetizations in spin-Hall oscillators via spin current injection}
\author{Tomohiro~Taniguchi
        \\
        National Institute of Advanced Industrial Science and Technology (AIST), 
              Spintronics Research Center, 
              Tsukuba, Ibaraki 305-8568, Japan 
}

\maketitle

\begin{abstract}

An array of spin torque oscillators (STOs) for practical applications such as pattern recognition was recently proposed, 
where several STOs are connected by a common nonmagnet. 
In this structure, in addition to the electric and/or magnetic interactions proposed in previous works, 
the STOs are spontaneously coupled to each other through the nonmagnetic connector, due to the injection of spin current. 
Solving the Landau-Lifshitz-Gilbert equation numerically for such system consisting of three STOs driven by the spin Hall effect, 
it is found that both in-phase and antiphase synchronization of the STOs can be achieved 
by adjusting the current density and appropriate distance between the oscillators. 

\end{abstract}


\IEEEpeerreviewmaketitle


\section{Introduction}
\label{sec:Introduction}



\IEEEPARstart{A}{n} excitation of a mutually coupled motion of the magnetizations in nanostructured ferromagnets, such as synchronization between spin torque oscillators (STOs) [1-15], 
has attracted much attention from the viewpoints of both fundamental physics and practical applications 
such as phased arrays and brain-inspired computing [16,17]. 
The mechanism of the synchronization in the previous works was based on 
the electric and/or magnetic interactions among STOs, 
such as spin wave propagation, current injection, microwave application, and dipole interaction. 


Spintronics devices have another possibility to excite a coupled dynamics of magnetizations by an injection of spin current. 
For example, the coupled motion of two ferromagnets in ferromagnetic resonance through spin pumping was studied previously [18-20]. 
Recently, we studied a synchronization of self-oscillations between STOs by the injection of spin current [21]. 
The system we considered was similar to an array of STOs proposed by Kudo and Morie for pattern recognition [17], 
where several STOs driven by the spin Hall effect [22-27] are connected by a common nonmagnetic electrode. 
Note that a self-oscillation in each STO is excited when a spin current is injected from a nonmagnetic heavy metal into the free layer of the STO. 
We noticed that the spin current simultaneously creates spin accumulation inside the free layer. 
When the free layers of the STOs are connected by a nonmagnet having a long spin diffusion length, 
another spin current flows in the connector, in accordance with the gradient of the spin accumulation. 
This additional spin current excites additional spin torques on the magnetizations, and leads to a coupled motion of the magnetizations. 
Considering two STOs, we showed that this type of coupling results in an antiphase synchronization of the magnetizations. 
The result indicates a possibility to excite a spontaneous synchronization between STOs without using electric and/or magnetic interactions. 
It is of interest accordingly to extend the system to a large number of STOs. 


In this paper, theoretical investigation is given for the phase dynamics between three STOs driven by the spin Hall effect. 
It is found that two of three STOs show phase synchronizations, whereas the other STO shows an oscillation with a different frequency. 
An antiphase synchronization between two STOs is found for a relatively large-coupling case. 
For a relatively weak coupling case, on the other hand, 
the antiphase synchronization appears for a small current region, 
whereas the phase difference becomes an in-phase for a large current region. 



\section{System description}
\label{sec:System description}



\begin{figure}
\centerline{\includegraphics[width=1.0\columnwidth]{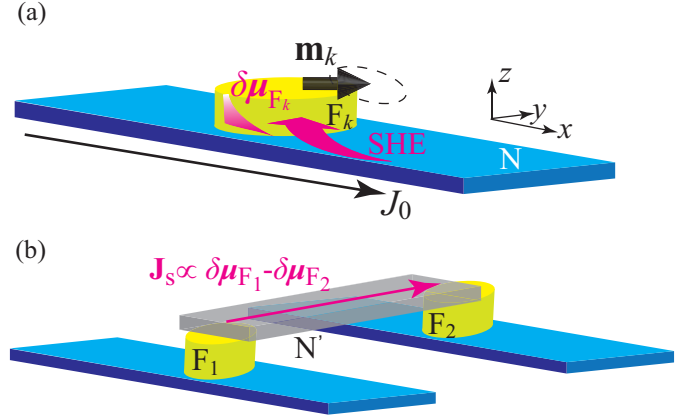}}
\caption{
         (a) Schematic view of spin-Hall oscillator with the spin Hall effect (SHE). 
             The electric current density $J_{0}$ in the bottom nonmagnet (N) flowing in the $x$-direction is converted to a spin current moving to the $z$-direction by the SHE, 
             and excites an oscillation of the magnetization $\mathbf{m}_{k}$ in the $k$-th ferromagnet F${}_{k}$. 
             The spin current also creates the spin accumulation $\delta\bm{\mu}_{{\rm F}_{k}}$. 
         (b) Two spin-Hall oscillators are connected by a nonmagnet N${}^{\prime}$ on the tops of the ferromagnets. 
             In this case, another spin current density $\mathbf{J}_{\rm s}$ proportional to the difference of the spin accumulation flows in the connector. 
         }
\label{fig:fig1}
\end{figure}



The basic idea of the coupling mechanism between $N$ STOs is as follow, where $N$ is the number of the oscillators. 
Each oscillator consists of a ferromagnet and a nonmagnetic heavy metal placed at the bottom. 
We use suffixes such as $k,k^{\prime}=1,2,...,N$ to distinguish the ferromagnets F${}_{k}$. 
Electric current densities $J_{0}$ along the $x$-direction are applied to all of the bottom nonmagnet. 
The ferromagnet is placed onto the nonmagnet along the $z$-direction, as shown in Fig. \ref{fig:fig1}(a). 
According to the experiment [28], 
we assume that the internal magnetic field $\mathbf{H}_{k}$ of the ferromagnet F${}_{k}$ consists of an in-plane anisotropy field $H_{\rm K}$ along the $y$-direction 
and a demagnetization field $-4\pi M$ along the $z$-direction as 
\begin{equation}
  \mathbf{H}_{k}
  =
  H_{\rm K}
  m_{ky}
  \mathbf{e}_{y}
  -
  4\pi M 
  m_{kz}
  \mathbf{e}_{z}, 
  \label{eq:field}
\end{equation}
where $\mathbf{m}_{k}=(m_{kx},m_{ky},m_{kz})$ is the unit vector pointing in the magnetization direction of the F${}_{k}$ layer. 
The magnetic energy density of the ferromagnet is given by $E_{k}=-M \int d \mathbf{m}_{k}\cdot\mathbf{H}_{k}=-(MH_{\rm K}/2)m_{ky}^{2}+2\pi M^{2} m_{kz}^{2}$. 
The energetically stable states correspond to $\mathbf{m}_{k}=\pm\mathbf{e}_{y}$. 


The spin Hall effect in the bottom nonmagnet injects pure spin current having the spin polarization along the $y$-direction into the F${}_{k}$ layer, 
and excites the spin torque 
\begin{equation}
  \mathbf{T}_{k}
  =
  -\frac{\gamma \hbar \vartheta_{\rm R} J_{0}}{2eMd}
  \mathbf{m}_{k}
  \times
  \left(
    \mathbf{e}_{y}
    \times
    \mathbf{m}_{k}
  \right), 
  \label{eq:SOT}
\end{equation}
where $\gamma$, $M$, and $d$ are the gyromagnetic ratio, saturation magnetization, and thickness of the ferromagnet, respectively. 
An effective spin Hall angle, including the interface mixing conductance, is denoted as $\vartheta_{\rm R}$ [13]. 
The spin torque given by  Eq. (\ref{eq:SOT}) induces a self-oscillation of the magnetization $\mathbf{m}_{k}$ around the $y$-direction [28]. 
We note that the pure spin current generated from the bottom nonmagnet 
simultaneously creates the spin accumulation in the F${}_{k}$ layer, 
which obeys the diffusion equation [25,29] and is given by [30] 
\begin{equation}
  \delta
  \bm{\mu}_{{\rm F}_{k}}(z)
  =
  e 
  \vartheta^{*}
  \lambda_{\rm F}
  E_{x}
  m_{ky}
  \cosh
  \left(
    \frac{z-d}{\lambda_{\rm F}}
  \right)
  \mathbf{m}_{k}. 
  \label{eq:spin_accumulation}
\end{equation}
Here, $\vartheta^{*}=\vartheta\{ \sigma_{\rm N} g^{*} \tanh[d_{\rm N}/(2\lambda_{\rm N})]/[(1-\beta^{2})\sigma_{\rm F}g_{\rm N}\sinh(d/\lambda_{\rm F})]\}$ [21] 
depends on the conductivity $\sigma$ and spin-diffusion length $\lambda$, 
where we use the suffixes F and N to distinguish the quantities related to the ferromagnet and nonmagnet, respectively. 
The thickness of the bottom nonmagnet is $d_{\rm N}$. 
The pure spin Hall angle in the nonmagnet and the spin polarization of the conductivity in the ferromagnet are $\vartheta$ and $\beta$, respectively. 
The quantity $g^{*}$ is related to the F/N interface resistance, whereas $g_{\rm N}/S$ with the F/N cross section area $S$ is defined as $g_{\rm N}/S=h \sigma_{\rm N}/(2e^{2}\lambda_{\rm N})$ [30]. 
The origin of the $z$ axis locates at the F/N interface. 


Now let us consider a coupling between the ferromagnets. 
We note that the spin accumulation given by Eq. (\ref{eq:spin_accumulation}) depends on the magnetization direction. 
Therefore, even when all the ferromagnets have the same magnetic properties and are under the effect of the same current densities, 
the spin accumulations in the ferromagnets are different when the magnetizations point to different directions. 
When the top surfaces of two ferromagnets, F${}_{k}$ and F${}_{k^{\prime}}$, are connected by an additional nonmagnet N${}^{\prime}$, as shown in Fig. \ref{fig:fig1}(b), 
another spin current flows in the connector according to the gradient of the spin accumulation $\delta\bm{\mu}_{{\rm F}_{k}}$. 
When the spin-diffusion length of the connector is sufficiently longer than its dimensional length $L$, 
the spin current in the top connector flowing from the F${}_{k}$ to F${}_{k^{\prime}}$ layer is given by 
\begin{equation}
  \mathbf{J}_{\rm s}^{{\rm F}_{k} \to {\rm F}_{k^{\prime}}}
  \simeq
  \frac{\hbar \sigma_{\rm N^{\prime}}}{2e^{2}L}
  \left[
    \delta
    \bm{\mu}_{{\rm F}_{k}}
    (z=d)
    -
    \delta
    \bm{\mu}_{{\rm F}_{k^{\prime}}}
    (z=d)
  \right].
  \label{eq:coupling_spin_current}
\end{equation}
where $\sigma_{\rm N^{\prime}}$ is the conductivity of the top connector. 
The emission of the spin current given by Eq. (\ref{eq:coupling_spin_current}) from the F${}_{k}$/N${}^{\prime}$ interface results in 
an excitation of an additional spin torque acting on $\mathbf{m}_{k}$ given by 
\begin{equation}
  \mathbf{T}_{kk^{\prime}}
  =
  -\frac{\gamma \hbar \tilde{\vartheta} J_{0}}{2eMd}
  m_{k^{\prime}y}
  \mathbf{m}_{k}
  \times
  \left(
    \mathbf{m}_{k^{\prime}}
    \times
    \mathbf{m}_{k}
  \right), 
  \label{eq:coupling_torque}
\end{equation}
where we use Eqs. (\ref{eq:spin_accumulation}) and (\ref{eq:coupling_spin_current}). 
We introduce $\tilde{\vartheta}$ as
\begin{equation}
  \tilde{\vartheta}
  =
  \vartheta^{*}
  \frac{\sigma_{\rm N^{\prime}} \lambda_{\rm F}}{\sigma_{\rm N} L}. 
  \label{eq:coupling_strength}
\end{equation}
Using Eqs. (\ref{eq:SOT}) and (\ref{eq:coupling_torque}), 
the Landau-Lifshitz-Gilbert equation of the magnetization is given by 
\begin{equation}
  \frac{d \mathbf{m}_{k}}{dt}
  =
  -\gamma
  \mathbf{m}_{k}
  \times
  \mathbf{H}_{k}
  +
  \alpha
  \mathbf{m}_{k}
  \times
  \frac{d \mathbf{m}_{k}}{dt}
  +
  \mathbf{T}_{k} 
  +
  \sum_{k^{\prime} \neq k}
  \mathbf{T}_{kk^{\prime}},
  \label{eq:LLG}
\end{equation}
where $\alpha$ is the Gilbert damping constant. 
We should note that the coupling torque $\mathbf{T}_{kk^{\prime}}$ in Eq. (\ref{eq:LLG}) results in a coupled motion of the magnetizations 
because it depends on the magnetization directions $\mathbf{m}_{k^{\prime}}$ in the other ferromagnets. 
For a system consisting of two ($N=2$) STOs, it was shown that this coupling torque leads to an antiphase synchronization of the magnetizations [21]. 
However, a coupled dynamics between STOs for $N \ge 3$ has not been investigated yet. 



\section{Synchronization of three STOs}
\label{sec:Synchronization of three STOs}

\begin{figure*}
\centerline{\includegraphics[width=2.0\columnwidth]{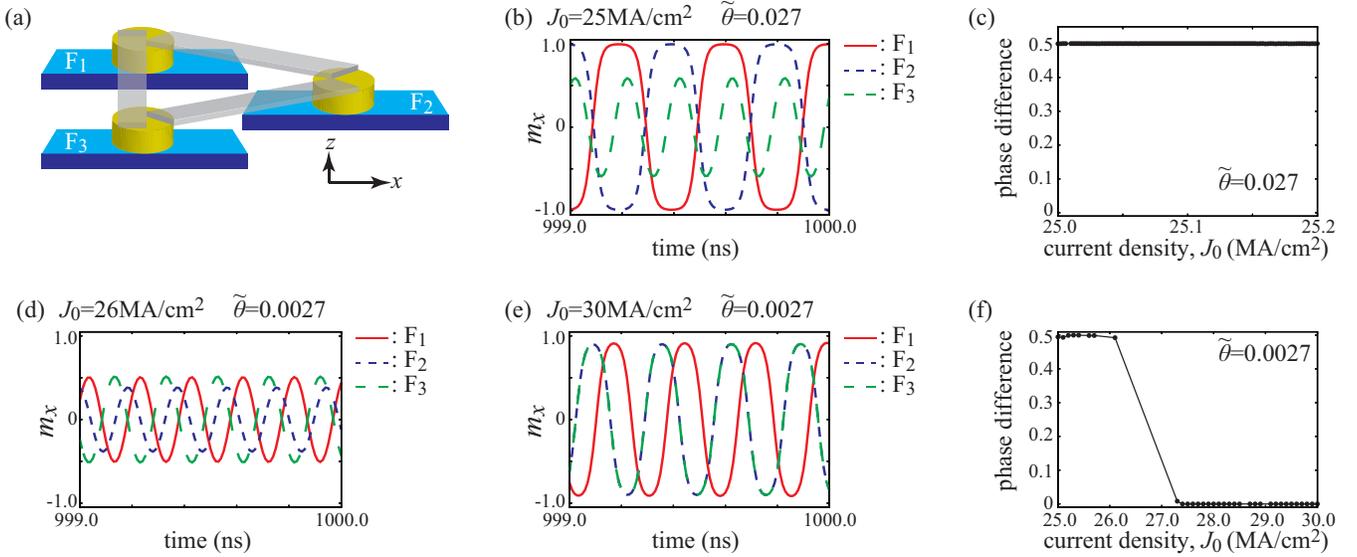}}
\caption{
        (a) Schematic view of a system consisting of three STOs. 
            Each ferromagnet is located at the vertex of an equilateral triangle to make the coupling strength between STOs identical. 
        (b) Oscillations of $m_{kx}$ in a steady state for a relatively large coupling strength, $\tilde{\vartheta}=0.027$ [21],
            where $J_{0}=25.0$ MA/cm${}^{2}$. 
            The red solid, blue dotted, and green dashed lines correspond to $m_{1x}$, $m_{2x}$, and $m_{3x}$, respectively. 
        (c) Current dependence of the phase difference between two STOs for the strong coupling strength. 
            The values $0$ and $0.5$ in the vertical axis correspond to the in-phase and antiphase, respectively. 
        (d), (e) Oscillations of $m_{kx}$ for a relatively weak coupling strength, $\tilde{\vartheta}=0.0027$, where $J_{0}=26.0$ and $30.0$ MA/cm${}^{2}$. 
        (f) Current dependence of the phase difference between two STOs for the weak coupling strength. 
         }
\label{fig:fig2}
\end{figure*}

We study the coupled motion of the magnetizations by solving Eq. (\ref{eq:LLG}) numerically. 
It has been revealed in the field of nonlinear science that 
even the behavior of a small number of identical oscillators is rather complex [31]. 
For example, for three oscillators arranged in a ring coupled through electric interaction, 
three stable synchronous states are possible, depending on the coupling strength [32].  
In our case, we note that the coupling strength $\tilde{\vartheta}$ depends on the distance $L$ between the ferromagnets. 
This fact means that the maximum number of the oscillators to connect all of them by the same coupling strength in two-dimensional space is three. 
Therefore, we consider the case of $N=3$, 
and assume that each ferromagnet is located at the vertex of an equilateral triangle. 
Figure \ref{fig:fig2}(a) shows a possible alignment of the STOs.


The material parameters are derived from recent experiments on the spin Hall magnetoresistance in W/CoFeB metallic bilayer [33] and first-principles calculations [34] as 
$M=1500$ emu/c.c., $H_{\rm K}=200$ Oe, $\gamma=1.764 \times 10^{7}$ rad/(Oe s), $\alpha=0.005$, $d=2$ nm, and $\vartheta_{\rm R}=0.167$. 
The value of the coupling strength, $\tilde{\vartheta}$, for $L=100$ nm was estimated to be $\tilde{\vartheta}=0.027$ 
by assuming that N${}^{\prime}$ consists of Cu [21]. 
In this paper, we also study the case of a weak coupling, $\tilde{\vartheta}=0.0027$. 
We note that, in the absence of the coupling, the self-oscillation in an STO is excited 
when the current density $J_{0}$ is in the range of $J_{\rm c}<|J_{0}|<J^{*}$ [35], where 
\begin{equation}
  J_{\rm c}
  =
  \frac{2\alpha eMd}{\hbar \vartheta_{\rm R}}
  \left(
    H_{\rm K}
    +
    4\pi M 
  \right),
  \label{eq:Jc}
\end{equation}
\begin{equation}
  J^{*}
  =
  \frac{4 \alpha eMd}{\pi \hbar \vartheta_{\rm R}}
  \sqrt{
    4\pi M 
    \left(
      H_{\rm K}
      +
      4\pi M 
    \right)
  }.
  \label{eq:J_star}
\end{equation}
The critical current density $J_{\rm c}$ is the minimum current density necessary to destabilize the magnetization staying near the easy axis and excites self-oscillation. 
On the other hand, $J^{*}$ is the switching current density to reverse the magnetization direction between two stable states. 
The values of $J_{\rm c}$ and $J^{*}$ are 26 and 33 MA/cm${}^{2}$, respectively. 
We, however, emphasize that these values are defined for a single STO. 
The current densities determining the oscillation range are mainly determined by the effective spin Hall angle $\theta_{\rm R}$ 
but are slightly affected by the coupling constant $\tilde{\theta}$, as shown below. 


Figure \ref{fig:fig2}(b) shows an example of the coupled motion of the magnetizations for a relatively strong coupling strength $\tilde{\vartheta}=0.027$ and $J_{0}=25$ MA/cm${}^{2}$. 
Here, the oscillations of the $x$-components of $\mathbf{m}_{k}$ ($k=1,2,3$) in a steady state are shown by different lines. 
We see that the magnetizations in two STOs show the antiphase synchronization, whereas the magnetization in the other STO shows a self-oscillation with a different frequency. 
For example, in Fig. \ref{fig:fig2}(b), 
the magnetizations in F${}_{1}$ (red solid) and F${}_{2}$ (blue dotted) layers are antiphase-coupled. 
On the other hand, the magnetization in F${}_{3}$ (green dashed) layer shows an oscillation with a frequency which is higher than that of F${}_{1}$ and F${}_{2}$ layers. 
We note that which of three STOs are coupled depend on the initial conditions of the magnetizations. 
The phase difference between the coupled STOs is, however, independent from which of STOs are coupled. 
It should be emphasized that the antiphase synchronization of two STOs appear even for different values of the current. 
Figure \ref{fig:fig2}(c) summarizes the current dependence of the phase difference between coupled STOs,  
where $0$ and $0.5$ in the vertical axis correspond to the in-phase and antiphase, respectively [13]. 
It shows that the antiphase synchronization between two STOs appears for the current range of $25.0 \le J_{0} \le 25.2$ MA/cm${}^{2}$. 
We find that the magnetizations relax to the stable state $\mathbf{m}_{k}=+\mathbf{e}_{y}$ near the initial state for $J_{0}<25.0$ MA/cm${}^{2}$, 
whereas they switch to the other stable state $\mathbf{m}_{k}=-\mathbf{e}_{y}$ for $J_{0}>25.3$ MA/cm${}^{2}$. 
Also it should be noticed here that the current range of the self-oscillation is significantly suppressed by the coupling torque. 


We also investigate a coupled motion of the magnetizations for a relatively weak coupling strength, $\tilde{\vartheta}=0.0027$. 
Be reminded that the coupling strength can be adjusted by changing the distance between the STOs, as can be seen from Eq. (\ref{eq:coupling_strength}). 
Figures \ref{fig:fig2}(d) and \ref{fig:fig2}(e) show examples of the magnetization oscillations for $J_{0}=26.0$ and $30.0$ MA/cm${}^{2}$, respectively. 
For $J_{0}=26.0$ MA/cm${}^{2}$, two magnetizations show an antiphase synchronization, whereas the other magnetization oscillate with a different frequency. 
This behavior is similar to the result shown in Fig. \ref{fig:fig2}(b). 
On the other hand, for $J_{0}=30.0$ MA/cm${}^{2}$, two magnetizations show an in-phase synchronization, 
i.e., the magnetizations in F${}_{2}$ and F${}_{3}$ layers oscillate with the same phases. 
Figure \ref{fig:fig2}(f) summarizes the current dependence of the phase difference between coupled STOs, 
indicating that the antiphase synchronization is stabilized in a relatively small current region 
whereas the in-phase synchronization appears in a relatively large current region. 


An in-phase synchronization between oscillators is useful to enhance the emission power from devices such as microwave generator and magnetic sensor. 
On the other hand, an antiphase synchronization, or more generally, out-of-phase synchronization, can be used in practical devices 
such as phased array and pattern recognition [17,36-40]. 
Therefore, a precise control of the phases between STOs is of interest in applied physics. 
The result shown in Fig. \ref{fig:fig2}(f) indicates a possibility to control 
the phase difference between the magnetizations in a spin Hall geometry 
by adjusting the current density and choosing an appropriate distance between the STOs. 



\section{Summary}
\label{sec:Summary}

In conclusion, theoretical investigation is carried out in a coupled motion of three magnetizations in a spin Hall geometry. 
The ferromagnets are coupled to each other through the injection of spin current by connecting the top surfaces of the free layers with nonmagnets having long spin diffusion lengths. 
For a relatively strong coupling case, two magnetizations showed an antiphase synchronization whereas the other magnetization oscillated with a different frequency. 
The current range of the self-oscillation was significantly suppressed compared with the case of free running. 
For a relatively weak coupling case, on the other hand, the phase difference between two STOs depended on the current magnitude. 
The antiphase synchronization appeared when the current was small, 
whereas an in-phase synchronization was found in the large current region. 
The results indicate a possibility to achieve a precise control of the phases in the STOs 
by adjusting the current density and choosing an appropriate distance between them. 


\section*{Acknowledgement}

This work was supported by JSPS KAKENHI Grant-in-Aid for Young Scientists (B) 16K17486. 
The author is grateful grateful to Y. Kawamura, S. Tsunegi, T. Yorozu, and H. Kubota for valuable discussions.
The author is also thankful to S. Iba, A. Spiesser, H. Maehara, and A. Emura for their support and encouragement.





\end{document}